\documentclass[aoas,MSNbibl,nameyear,seceqn,dvips]{arximspdf}
\usepackage{multirow}
\usepackage{dcolumn}
\usepackage{graphicx}

\doi{10.1214/12-AOAS545} 
\volume{6}
\issue{3}
\pubyear{2012}
\firstpage{1134}
\lastpage{1161}

\makeatletter

\let\sv@tabnotetext\tabnotetext
\let\sv@tabnotemark@fmt\tabnotemark@fmt
\long\def\legend#1{{\let\tabnote@indent\leavevmode\sv@tabnotetext[]{}{#1}}}

\newcolumntype{d}[1]{D{.}{.}{#1}}

\newcommand{\blm}{\bolds}
\newcommand{\tr}{\operatorname{tr}}
\newcommand{\AM}{\operatorname{AM}}

\renewcommand{\epsilon}{\varepsilon}

\makeatother

\begin{document}
\begin{frontmatter}

\title{Gene-centric gene--gene interaction: A~model-based~kernel
machine method\thanksref{T1}}
\runtitle{Gene-centric gene--gene interaction}

\thankstext{T1}{Supported in part by NSF Grants DMS-0707031,
MCB-1121650 and by the Intramural Research Program of the
\textit{Eunice Kennedy Shriver} National Institute of Child Health and
Human Development, NIH, DHHS.}

\begin{aug}
\author[A]{\fnms{Shaoyu} \snm{Li}\ead[label=e1]{Shaoyu.Li@stjude.org}}
\and
\author[B]{\fnms{Yuehua} \snm{Cui}\corref{}\ead[label=e2]{cui@stt.msu.edu}}
\runauthor{S. Li and Y. Cui}
\affiliation{Michigan State University and St. Jude Children's Research Hospital}
\address[A]{Department of Statistics and Probability\\
Michigan State University\\
East Lansing, Michigan 48824\\
USA\\
and\\
Department of Biostatistics\\
St. Jude Children's Research Hospital\\
Memphis, Tennessee 38105\\
USA\\
\printead{e1}}
\address[B]{Department of Statistics and Probability\\
Michigan State University\\
A432 Wells Hall \\
East Lansing, Michigan 48824\\
USA\\
\printead{e2}}
\end{aug}

\received{\smonth{10} \syear{2011}}
\revised{\smonth{1} \syear{2012}}

%
\begin{abstract}
Much of the natural variation for a complex trait can be explained by
variation in DNA sequence levels. As part of sequence variation,
gene--gene interaction has been ubiquitously observed in nature, where
its role in shaping the development of an organism has been broadly
recognized. The identification of interactions between genetic factors
has been progressively pursued via statistical or machine learning
approaches. A large body of currently adopted methods, either
parametrically or nonparametrically, predominantly focus on pairwise
single marker interaction analysis. As genes are the functional units
in living organisms, analysis by focusing on a~gene as a system could
potentially yield more biologically meaningful results. In this work,
we conceptually propose a gene-centric framework for genome-wide
gene--gene interaction detection. We treat each gene as a~testing unit
and derive a model-based kernel machine method for two-dimensional
genome-wide scanning of gene--gene interactions. In addition to the
biological advantage, our method is statistically appealing because it
reduces the number of hypotheses tested in a genome-wide scan.
Extensive simulation studies are conducted to evaluate the performance
of the method. The utility of the method is further demonstrated with
applications to two real data sets. Our method provides a conceptual
framework for the identification of gene--gene interactions which could
shed novel light on the etiology of complex diseases.
\end{abstract}

%

\end{frontmatter}

\section{Introduction}
Accumulative evidence shows that much of the genetic variation
for a complex trait can be explained by the joint function of
multiple genetic factors as well as environmental contributions.
Searching for these contributing genetic factors and further
characterizing their effect sizes is one of the primary goals and
challenges for modern genetics. The recent breakthroughs in
high-throughput genotyping technologies and the completion of the
International HapMap project provide unprecedented opportunities to
characterize the genetic machinery of
living organisms. Genetic association analyses focusing on single
nucleotide polymorphisms (SNPs) or haplotypes have led to the
identification of many novel genetic determinants of complex disease traits.
However, despite enormous success in genome-wide association
studies, single SNP or haplotype based studies still suffer from low
replication rates because of the infeasibility of dealing with the
complex patterns of association, for example, genetic heterogeneity,
gene--gene interaction and gene-environment interaction.
Many of the genetic components of many diseases remain
unaccounted for and only a small proportion of the heritability has
been explained.

It is commonly recognized that genes do not function alone, rather they
interact constantly with each other. Gene--gene interactions have been
broadly considered as important contributors to the unexplained
heritability of complex traits [\citet{ThoMooHai04},
\citet{Mah08}, \citet{MooWil09}, \citet{Eicetal10}].
Current methods for gene--gene interactions are mostly focused on
single locus interactions, using either parametric methods such as the
regression-based tests of interaction [\citet{PieWeiTay94}] and
Bayesian epistasis mapping [\citet{ZhaLiu07}], nonparametric
methods such as the entropy-based approaches [\citet{Kanetal08}],
or data mining methods such as the multifactor dimensionality reduction
(MDR) [\citet{Ritetal01}] and random forests [\citet{Bre01}].
Methods based on interaction of haplotypes have also been developed
[e.g., \citet{Lietal10}, \citet{LiZhaYi11}]. Due to the issue
of haplotype phase-ambiguity, however, haplotype-based interaction
analysis is limited to small sized haplotypes. Extension to large sized
haplotype interaction is computationally challenging. For a
comprehensive review of statistical methods developed for detecting
gene--gene interactions, readers are referred to \citet{Cor09}.

A number of research reports have argued the relative merit of
gene-based association analysis [e.g., \citet{NeaSha04},
\citet{JorWit06}, \citet{Cuietal08}, \citet{Maetal10}].
\citet{NeaSha04} argued that a gene-based approach, in which all
variants within a putative gene are considered jointly, have relative
advantages over single SNP or haplotype analysis. As genes are the
functional units in a human genome, variants in genes should have high
probability of being functionally more important than those that occur
outside of a gene [\citet{JorWit06}]. Because of this
characteristic, gene-based association analysis would provide more
biologically interpretable results than the single-SNP or haplotype
based analysis. Moreover, when multiple variants within a gene function
in a complicated manner, the gene-based association test can gain
additional power by capturing the joint function of multiple variants
simultaneously compared to a single SNP analysis
[\citet{Cuietal08}, \citet{BuiMar}]. In addition, a
gene-based analysis is statistically appealing. By considering multiple
SNP markers within a gene as testing units, one can reduce the number
of tests, hence releasing the multiple testing burden and improving
association test power.

The relative advantage of gene-centric analysis motivates us to
consider genes as modeling units to identify interactions in a gene
level. It is our expectation that the
identification of genetic interactions in a gene level should carry the
same benefits and gains as it does
with gene-based association analysis. We therefore propose to jointly
model the
genetic variation of SNPs within a gene, then further test the
interaction in a gene level rather than in a single SNP level. This
conceptual definition and modeling of gene-centric gene--gene (denoted
as 3G) interaction would change the traditional paradigm of gene--gene
interaction analysis and help us gain novel insight into the genetic
etiology of complex diseases. In addition to its biological merits, by
focusing on genes as testing units, the number of
pairwise interaction tests can be dramatically reduced compared to a
single SNP-based pairwise interaction analysis. Thus, the 3G
interaction analysis is also statistically appealing.

Following the definition of the gene-centric interaction, we propose
a~model-based kernel machine method to identify significant gene--gene
interactions under the
proposed 3G analysis framework. Kernel-based methods have been
proposed to evaluate association of genetic variants with complex
traits in the past decades [e.g., \citet{Tzeetal03},
\citet{Schetal05},
\citet{WesSch06}, Schaid (\citeyear{Sch10N1}, \citeyear
{Sch10N2})]. A general kernel machine
method can account for complex nonlinear SNP effects within
a~genetic feature (e.g., a~gene or a pathway) by using an appropriately
selected kernel function. Generally speaking, a kernel function
captures the pairwise genomic similarity between individuals for
variants within an appropriately defined feature [\citet{Sch10N1}].
The application of kernel-based methods in genetic association
analysis has been reported in the literature [e.g.,
\citet{Schetal05},
\citet{Kweetal08}, \citet{Wuetal10}], but none of them
considers gene--gene interactions. In this work, we propose a general
3G interaction framework by applying the smoothing-spline ANOVA
model [\citet{Wah90}] to model gene--gene interactions. The proposed
method, termed {G}ene-centric {G}ene--{G}ene
interaction with {S}moothing-s{P}line {A}NOVA
Model (3G-SPA), is implemented through a two-step procedure:
(1) an exhaustive two-dimensional genome-wide search for any genetic
effects; and (2) assessment of significance of interactions for the
identified gene pairs.

The rest of the paper is organized as follows. In Section
\ref{methods} we describe the detailed model derivation of our
method. We propose two score statistics for testing the overall
genetic effect and the interaction effect. To
evaluate the performance of the proposed method,\vadjust{\goodbreak} Monte Carlo
simulations are performed in Section~\ref{simulation}. The utility
of the method is demonstrated by two real data analyses in Section
\ref{case}, followed by discussion in Section~\ref{discussion}.\vspace*{-2pt}

\section{Statistical methods}\label{methods}\vspace*{-2pt}
\subsection{Smoothing spline-ANOVA model}

We assume $n$ unrelated individuals sampled
from a population, each of which possesses a measurement for a
quantitative disease trait of interest. The quantitative measurements
of $n$ individuals are denoted as ${\mathbf y}=(y_{1}, y_{2},\ldots,
y_{n})^{T}$. Traditional approaches for detecting gene--gene
interactions, such as MDR or regression type analysis, identify
SNP-SNP interactions. In this work, we focus our attention to
pairwise gene--gene interactions by considering each gene as a unit.
Consider two genes, denoted as $G_1$ and $G_2$, with $L_{1}$ and
$L_{2}$ SNP markers, respectively.
Let ${\mathbf x}_{i}=(x_{i, 1},\ldots, x_{i,L})$ be an $1\times L$
genotype vector of the gene pair for subject~$i$, where
$L=L_{1}+L_{2}$ is the total number of SNP markers in the two genes.
We model the relationship between the genotypes of the gene
pair~(${\mathbf x}_{i}$) and the phenotype $y_{i}$ by the following model:
%
%
\begin{equation}\label{model1}
y_{i}=m(\mathbf{x}_i)+\epsilon_{i},\qquad i=1, 2,\ldots, n,
\end{equation}
where $m$ is an unknown function and $\epsilon_{i}\sim\mathcal{N}(0,
\sigma_i^2)$ is a random subject-specific error term and independent of
${\mathbf x}_{i}$. Here $\sigma_i^2$ (\mbox{$=$}$\sigma^2$) is generally assumed to
be homogeneous.

\citet{Gu02} has discussed the ANOVA decomposition of multivariate
functions on generic domains of each single coordinate. Actually,
the decomposition can also be defined on nested domains (see
Appendix~\ref{Appendix1}). Following a~similar idea, the
genotype vector ${\mathbf x}_{i}$ is partitioned as ${\mathbf
x}_{i}=[{\mathbf x}_{i}^{(1)}, {\mathbf x}_{i}^{(2)}]$, where~${\mathbf x}_{i}^{(j)}$
represents the $L_j$ SNP predictors for gene $j$ ($j=1, 2$). Let a~product domain be
$\mathcal{X}=\mathcal{X}^{(1)}\otimes\mathcal{X}^{(2)}$ with ${\mathbf
x}^{(j)}\in\mathcal{X}^{(j)}$ and $A_{j}$ be an averaging operator
on $\mathcal{X}^{(j)}$, that averages out ${\mathbf x}_{i}^{(j)}$, $j=1,
2$. Then a function $m(\cdot)$ defined on the product domain has a
functional ANOVA decomposition as in the following:
%
%
\begin{eqnarray}\label{ANOVA-decomposite}
m&=&{\prod_{j=1}^2(I-A_j+A_j)}m\nonumber\\
&=&\{A_1A_2+(I-A_1)A_2+A_1(I-A_2)+(I-A_1)(I-A_2)\}m\\
&=&\mu+m_1+m_2+m_{12},\nonumber
\end{eqnarray}
where $\mu$ is the overall mean, $m_1, m_2$ are the main effects of the
two genes and $m_{12}$ describes the interaction effect between
them (see Appendix~\ref{Appendix1} for more details).\vspace*{-2pt}

\subsection{Reproducing kernel Hilbert space and the dual representation}
Based on the ANOVA decomposition, a reproducing kernel Hilbert
space (RKHS)~$\mathcal{H}$ of functions on $\mathcal{X}$ can be
constructed\vadjust{\goodbreak} [\citet{GuWah93} and \citet{Wahetal95}].
Let~$\mathcal{H}^{(j)}$ be an RKHS of functions on
$\mathcal{X}^{(j)}$, $j=1, 2$, and $\mathbf{1}^{(j)}$ be a space
of constant functions on $\mathcal{X}^{(j)}$, then
%
%
\begin{eqnarray}\label{RKHS-ANOVA}
\mathcal{H}&=&\prod_{j=1}^{2}\bigl(\mathbf{1}^{(j)}\oplus{\mathcal
{H}^{(j)}}\bigr)\nonumber\\
&=&[\mathbf{1}]\oplus\bigl[\mathcal{H}^{(1)}\otimes\mathbf{1}^{(2)}\bigr]\oplus
\bigl[\mathbf{1}^{(1)}\otimes{\mathcal{H}^{(2)}}\bigr]\oplus\bigl(\mathcal
{H}^{(1)}\otimes{\mathcal{H}^{(2)}}\bigr)\\
&=&[\mathbf{1}]\oplus\mathcal{H}^{1}\oplus\mathcal{H}^{2}\oplus
\mathcal{H}^{3},\nonumber
\end{eqnarray}
where $\oplus$ refers to direct sum and $\otimes$ refers to
tensor product. Equation (\ref{RKHS-ANOVA}) provides an orthogonal
decomposition of the entire functional space $\mathcal{H}$. So
$\mathcal{H}$ is a RKHS with the associated reproducing kernel as
the sum of the reproducing kernels of these component subspaces.
Each functional component in (\ref{ANOVA-decomposite}) lies in a
subspace in (\ref{RKHS-ANOVA}), and is estimated in the
corresponding RKHS. The identifiability of the components is assured
by side conditions: $\int_{\mathcal{X}^{(j)}}m_{j}({\mathbf
x}^{(j)})\,d\mu_{j}=0$, $j=1,2$.

We assume that function $m$ is a member of the RKHS
$\mathcal{H}$ and can be estimated as the minimizer of the
following penalized sum of squares:
%
%
\begin{equation}\label{PLike}
\mathcal{L}({\mathbf y}, m)=\sum_{i=1}^{n}\bigl(y_{i}-m({\mathbf
x}_{i})\bigr)^2+\lambda J(m),
\end{equation}
where $J(\cdot)$ is a roughness penalty. With the orthogonal decomposition
of space $\mathcal{H}$, the penalty function $J(\cdot)$ can be
decomposed such that equation~(\ref{PLike}) becomes
%
%
\begin{equation}\label{objective}
\mathcal{L}({\mathbf y}, m)=\sum_{i=1}^{n}\bigl(y_{i}-m({\mathbf
x}_{i})\bigr)^2+\sum
_{l=1}^{3}\lambda_{l}\|P^{l}m(\cdot)\|_{\mathcal{H}}^{2},
\end{equation}
where $P^{l}$ is the orthogonal projector in $\mathcal{H}$ onto
$\mathcal{H}^{l}$, and the $\lambda_l$'s are the tuning parameters which
balance the goodness of fit and complexity of the model. The
minimizer of the objective function (\ref{objective}) is known to have
a~representation
[\citet{Wah90}, Chapter 10] in terms of a~constant and the
associated reproducing kernels $\{k_{l}(s, t)\}$ of the
$\mathcal{H}^{l}, l=1, 2, 3$, that is,
%
%
\begin{eqnarray}
m({\mathbf x})&=&\mu+\sum_{i=1}^{n}c_{i}\sum_{l=1}^{3}\theta
_{l}k_{l}({\mathbf x}_{i}, {\mathbf x})\nonumber\\[-8pt]\\[-8pt]
&=&\mu+\sum_{l=1}^{3}K_{l}^{T}({\mathbf x})C_{l},
\nonumber
\end{eqnarray}
where $K_{l}^{T}({\mathbf x})=(k_{l}({\mathbf x}_{1}, {\mathbf
x}),\ldots,
k_{l}({\mathbf x}_{n}, {\mathbf x}))$, $C_{l}=(c_{1},\ldots,
c_{n})^{T}\theta_{l}$. Details on the choice of the reproducing
kernel functions corresponding to the three subspaces will be
discussed in a later section.

Substituting the representation of $m(\cdot)$ into (\ref
{objective}), we get
%
%
\begin{eqnarray}\quad
\mathcal{L}({\mathbf y}, m)&=&\sum_{i=1}^{n}\bigl(y_{i}-m({\mathbf
x}_{i})\bigr)^2+\sum
_{l=1}^3\lambda_{l}\|P^{l}m(\cdot)\|_{\mathcal
{H}}^{2}\nonumber\\
&=&\bigl({\mathbf y}-m({\mathbf X})\bigr)^{T}\bigl({\mathbf y}-m({\mathbf X})\bigr)+\sum
_{l=1}^{3}\lambda
_{l}C_{l}^{T}\mathbf{K}_{l}C_{l}\\
&=&\Biggl({\mathbf y}-\mu\mathbf{1}-\sum_{l=1}^3\mathbf
{K}_{l}C_{l}\Biggr)^{T}\Biggl({\mathbf y}-\mu\mathbf{1}-\sum_{l=1}^3\mathbf
{K}_{l}C_{l}\Biggr)+\sum_{l=1}^{3}\lambda
_{l}C_{l}^{T}\mathbf{K}_{l}C_{l},
\nonumber
\end{eqnarray}
where ${\mathbf X}=({\mathbf x}_{1}^{T},\ldots, {\mathbf
x}_{n}^{T})^{T}$ and
\[
\mathbf{K}_{l}=\left[
\matrix{
K_{l}^{T}({\mathbf x}_{1})\vspace*{1pt}\cr
K_{l}^{T}({\mathbf x}_{2})\cr
\vdots\vspace*{1pt}\cr
K_{l}^{T}({\mathbf x}_{n})}
\right].
\]
The gradients of $\mathcal{L}$ with respect to the coefficients
($\mu$, $C_{l}\dvtx l=1, 2, 3$) are
\[
\frac{\partial{\mathcal{L}}}{\partial{\mu}}=2\mathbf{1}^{T}\Biggl({\mathbf
y}-\mu
\mathbf{1}-\sum_{l=1}^{3}\mathbf{K}_{l}C_{l}\Biggr)
\]
and
\[
\frac{\partial{\mathcal{L}}}{\partial{C_{l}}}=2\Biggl\{\mathbf
{K}_{l}^{T}\Biggl({\mathbf y}-\mu\mathbf{1}-\sum_{l=1}^{3}\mathbf
{K}_{l}C_{l}\Biggr)+\lambda_{l}\mathbf{K}_{l}C_{l}\Biggr\}.
\]
Therefore, the first order condition is satisfied by the system
%
%
\begin{eqnarray}\label{sssystem}\qquad
&&
\left[
\matrix{
n & \mathbf{1}^{T}\mathbf{K}_{1}& \mathbf{1}^{T}\mathbf{K}_{2}& \mathbf
{1}^{T}\mathbf{K}_{3}\vspace*{2pt}\cr
\mathbf{K}_{1}^{T}\mathbf{1}& \mathbf{K}_{1}^{T}\mathbf{K}_{1}+\lambda
_{1}\mathbf{K}_{1}& \mathbf{K}_{1}^{T}\mathbf{K}_{2}& \mathbf
{K}_{1}^{T}\mathbf{K}_{3}\vspace*{2pt}\cr
\mathbf{K}_{2}^{T}\mathbf{1}& \mathbf{K}_{2}^{T}\mathbf{K}_{1}& \mathbf
{K}_{2}^{T}\mathbf{K}_{2}+\lambda_{2}\mathbf{K}_{2}& \mathbf
{K}_{2}^{T}\mathbf{K}_{3}\vspace*{2pt}\cr
\mathbf{K}_{3}^{T}\mathbf{1}& \mathbf{K}_{3}^{T}\mathbf{K}_{1}& \mathbf
{K}_{3}^{T}\mathbf{K}_{2}& \mathbf{K}_{3}^{T}\mathbf{K}_{3}+\lambda
_{3}\mathbf{K}_{3}}
\right]
\left[
\matrix{
\mu\cr C_{1} \cr C_{2}\cr C_{3}}
\right]\nonumber\\[-8pt]\\[-8pt]
&&\qquad=\left[
\matrix{
\mathbf{1}^{T}\vspace*{1pt}\cr
\mathbf{K}^{T}_{1}\vspace*{1pt}\cr
\mathbf{K}^{T}_{2}\vspace*{1pt}\cr
\mathbf{K}^{T}_{3}}
\right]{\mathbf y}.\nonumber
\end{eqnarray}

The connection between smoothing splines and the linear mixed effects
model has been previously established [\citet{Wah90},
\citet{Rob91}]. For the two-way ANOVA decomposition model
considered in this paper, we show that the first order system above is
equivalent to Henderson's normal equation of the following linear
mixed\vadjust{\goodbreak}
effects model (see Appendix~\ref{Equivalence} for details):
%
%
\begin{equation}\label{LMM}
\mathbf y=\mu\mathbf{1}+m_{1}+m_{2}+m_{12}+\epsilon,
\end{equation}
where $m_{1}, m_{2}, m_{12}$ are independent $n\times1$ vector of
random effects; $m_{1}\sim N(\mathbf{0}, \tau^{2}_{1}\mathbf{K}_{1})$,
$m_{2}\sim N(\mathbf{0}, \tau^{2}_{2}\mathbf{K}_{2})$, $m_{12}\sim
N(\mathbf{0}, \tau^{2}_{3}\mathbf{K}_{3})$, and $\epsilon\sim
N(\mathbf{0}, \sigma^2I)$ is independent of $m_{1}, m_{2}$ and
$m_{12}$. This connection indicates that the estimators of functions
$m_1, m_2, m_{12}$ are just the BLUPs of the linear mixed effects model
[see also \citet{LiuLinGho07}]. Tuning parameters $\lambda_{l},
l=1, 2, 3$, are functions of the variance components, which can be
estimated either by the maximum likelihood method or by the restricted
maximum likelihood (REML) method. Since the REML method produces
unbiased estimators for the variance components, we adopt REML
estimation in this work. The dual representation of the linear mixed
effects model obtained for the SS-ANOVA model makes it feasible to do
inferences about the main and interaction components under the mixed
effects model framework.

\subsection{Choice of kernel function for genotype similarity}
The choice of a~reproducing kernel is not arbitrary in the sense that
the kernel function must be nonnegative definite. By Theorem 2.3
[\citet{Gu02}], given a nonnegative definite function $k$ on
$\mathcal{X}$, we can construct a unique RKHS of real-valued functions
on $\mathcal{X}$ with $k$ as its reproducing kernel. In genetic
association studies, a kernel function captures the pairwise genomic
similarities across multiple SNPs in a gene. It projects the genotype
data from the original space, which can be high dimensional and
nonlinear, to a one-dimensional linear space. The allele matching (AM)
kernel is one of the most popularly used kernels for measuring genomic
similarity. This type of kernel measure has been used in linkage
analyses [\citet{WeeLan88}] and in association studies
[\citet{Tzeetal03}, \citet{Schetal05}, \citet{WesSch06},
\citet{Kweetal08}, \citet{Muketal10} and
\citet{Wuetal10}]. For a review of genomic similarity and kernel
methods, readers are referred to Schaid (\citeyear{Sch10N1},
\citeyear{Sch10N2}). With the notable strength that it does not require
knowledge of the risk allele for each SNP, the AM kernel is chosen as
the kernel function in this study. This similarity kernel counts the
number of matches among the four comparisons between two genotypes
$g_{i, s}$ (with two alleles $A$ and $B$) and $g_{j, s}$ (with two
alleles $C$ and $D$) of two individuals $i$ and $j$ at locus $s$, and
can be expressed as
\[
\AM(g_{i, s}=A/B, g_{j, s}=C/D)=I(A\equiv C)+I(A\equiv D)+I(B\equiv
C)+I(B\equiv D),
\]
where $I$ is the indicator function and ``$\equiv$'' means the two alleles
are identical-by-state (IBS). The kernel function based on this AM
similarity measure then takes the following form:
%
%
\begin{equation}
f(g_{i}, g_{j})=\frac{\sum_{s=1}^{S}\AM(g_{i,s}, g_{j,s})}{4S},
\end{equation}
where $S$ is the number of SNPs considered for each kernel function.\vadjust{\goodbreak}

To incorporate valuable SNP-specific information into analyses in order to
potentially improve performance, a weighted-AM kernel can be applied
which has the following form:
%
%
\begin{equation}
f(g_{i}, g_{j})=\frac{\sum_{s=1}^{S}w_{s}\AM(g_{i,s}, g_{j,s})}{4\sum
_{s=1}^{S}w_{s}},
\end{equation}
where $w_s$ is the weighting function which can be adopted to
incorporate prior knowledge in order to gain extra power. For example,
when a
study is trying to identify the effect of rare variants, the
weight function can be taken as the inverse of the minor allele
frequency to boost the signal for rare variants [\citet{Sch10N2}].
An example illustrating the calculation of the kernel matrix is
given in Appendix~\ref{example}.

We use the AM kernel as the reproducing kernel for the two
subspaces~$\mathcal{H}^{1}$ and $\mathcal{H}^{2}$ corresponding to
the main effects. Utilizing the fact that the reproducing kernel of
a tensor product of two reproducing kernel Hilbert spaces is the
product of the two reproducing kernels [\citet{Aro50}], the
associated reproducing kernel for $\mathcal{H}^{3}$ can be taken as
the product of the reproducing kernels of the two subspaces:
$\mathcal{H}^{1}$ and $\mathcal{H}^{2}$.

\subsection{Hypothesis testing}
\subsubsection{Testing overall genetic effect}
In a gene-based genetic association study, one is interested in
whether a gene as a system is associated with a~disease trait. In
the proposed 3G interaction study, we are interested in the
association of each gene with a quantitative trait as well as the
interaction between genes, if any. The analysis starts with a
two-dimensional pairwise search for gene pairs with overall
contribution to the phenotypic variation and then tests those
contributing gene pairs for interaction effect. Under the
SS-ANOVA framework, testing the overall genetic effect of a gene
pair is equivalent to testing $H_{0}\dvtx m_{1}=m_{2}=m_{12}=0$. Similarly,
testing for interaction effect can be
formulated as $H_{0}\dvtx m_{12}=0$.
With the linear mixed effects model representation, the
aforementioned two tests are equivalent to $H_{0}^{1}\dvtx
\tau^{2}_{1}=\tau^{2}_{2}=\tau^{2}_{3}=0$ and $H_{0}^{2}\dvtx
\tau^{2}_{3}=0$, respectively. Here, $\tau^{2}_{1}, \tau^{2}_{2},
\tau^{2}_{3}$ are the variance components in model
(\ref{LMM}).

A well-known issue in variance component analysis is that the
parameters under the null hypotheses are on the boundary of the
parameter space. Moreover, the kernel matrices $\mathbf{K}_{\ell}$'s
are not block-diagonal. Thus, the asymptotic distribution of
a likelihood ratio test (LRT) statistic does not follow a central
chi-square distribution under the null hypothesis. The mixture
chi-square distribution proposed by \citet{SelLia87} under
irregular conditions also does not apply in our case. In
this paper, we construct score test statistics based on the
restricted likelihood. Consider the linear mixed model in\vadjust{\goodbreak}
(\ref{LMM}), ${\mathbf y}\sim N(\mu\mathbf{1}, V(\beta))$. The
restricted log-likelihood function can be written as
\[
\ell_{R}\propto-\tfrac{1}{2}\ln(|V(\beta)|)-\tfrac{1}{2}\ln(|\mathbf
{1}^{T}V^{-1}(\beta)\mathbf{1}|)-\tfrac{1}{2}({\mathbf y}-\hat{\mu
}\mathbf
{1})^{T}V(\beta)^{-1}({\mathbf y}-\hat{\mu}\mathbf{1}),
\]
where $\beta=(\sigma^2, \tau^{2}_{1}, \tau^{2}_{2}, \tau^{2}_{3})^{T}$,
$V(\beta)=\sigma^2I+\tau^{2}_{1}\mathbf{K}_{1}+\tau^{2}_{2}\mathbf
{K}_{2}+\tau^{2}_{3}\mathbf{K}_{3}$.
The first order derivative of the restricted log-likelihood function
with respect to each variance component is
%
%
\begin{equation}\label{scorefunction}\quad
\frac{\partial\ell_{R}}{\partial\beta_{i}}=-\frac
{1}{2}\tr(RV_{i})+\frac{1}{2}({\mathbf y}-\hat{\mu}\mathbf
{1})^{T}V^{-1}(\beta)V_{i}V^{-1}(\beta)({\mathbf y}-\hat{\mu}\mathbf{1}),
\end{equation}
where $V_{i}=\frac{\partial V(\beta)}{\partial\beta_{i}}, i=1,\ldots
,4$, so that $V_{1}=I, V_{2}=\mathbf{K}_{1}, V_{3}=\mathbf{K}_{2},
V_{4}=\mathbf{K}_{3}$ and
$R=V^{-1}-V^{-1}\mathbf{1}(\mathbf{1}^{T}V^{-1}\mathbf{1})^{-1}\mathbf
{1}^{T}V^{-1}$.

The restricted score function under the null hypothesis $H_{0}^{1}\dvtx
\tau^{2}_{1}=\tau^{2}_{2}=\tau^{2}_{3}=0$ is given by
\[
\frac{\partial\ell_{R}}{\partial\beta_{i}}\bigg|_{\tau^{2}_{1}=\tau
^{2}_{2}=\tau^{2}_{3}=0}=-\frac{1}{2\sigma^{2}}\tr(P_{0}V_{i})+\frac
{1}{2\sigma^4}({\mathbf y}-\hat{\mu}\mathbf{1})^{T}V_{i}({\mathbf
y}-\hat{\mu
}\mathbf{1}),
\]
where $P_{0}=\mathbf{I}-\mathbf{1}(\mathbf{1}^{T}\mathbf{1})^{-1}\mathbf{1}^{T}$
is the projection matrix under the null. Thus, $H_{0}^{1}$ can be
tested using the following score statistic:
\[
S(\sigma^2)=\frac{1}{2\sigma^2}({\mathbf y}-\hat{\mu}_{0}\mathbf
{1})^{T}\sum
_{l=1}^{3}\mathbf{K}_{l}({\mathbf y}-\hat{\mu}_{0}\mathbf{1}),
\]
where $\hat{\mu}_{0}=(\mathbf{I}-P_{0})y$ is the MLE of $\mu$ under the
null. This leads to
\[
S(\sigma^2)=\frac{1}{2\sigma^2}{\mathbf y}^{T}P_{0}\sum_{l=1}^{3}\mathbf
{K}_{l}P_{0}{\mathbf y}.
\]
Denoting the true value of $\sigma^{2}$ under the null by $\sigma
_{0}^{2}$, $S(\sigma_{0}^2)$ is a quadratic form in ${\mathbf y}$.
Following \citet{LiuLinGho07}, we use the Satterthwaite method to
approximate the distribution of $S(\sigma_{0}^{2})$ by a scaled
chi-square distribution, that is, $S(\sigma_{0}^{2})\sim
a\chi^{2}_{g}$, where the scale parameter $a$ and the degrees of
freedom~$g$ can be estimated by the method of moments (MOM). By
equating the mean and variance of the test statistic
$S(\sigma_{0}^{2})$ with those of~$a\chi^{2}_{g}$, we have
\[
\cases{
\displaystyle \delta=E[S(\sigma_{0}^{2})]=\tr\Biggl(P_{0}\sum
_{i=1}^{3}\mathbf{K}_{i}\Biggr)\Big/2=E[a\chi^2_g]=ag,\vspace*{2pt}\cr
\displaystyle \nu=\operatorname{Var}[S(\sigma_{0}^{2})]=\tr\Biggl(\sum_{i=1}^{3}(P_{0}\mathbf{K}_{i})\sum
_{i=1}^{3}(P_{0}\mathbf{K}_{i})\Biggr)\Big/2=\operatorname{Var}[a\chi^2_g]=2a^2g.}
\]
Solving for the two equations leads to $\hat{a}=\nu/2\delta$ and $\hat
{g}=2\delta^2/\nu$.

In practice, we do not know the true value $\sigma_{0}^2$ and we
usually replace it by its MLE under the\vadjust{\goodbreak} null model, denoted by
$\hat{\sigma}_{0}^2$. The asymptotic distribution of $S(\hat{\sigma}
_{0}^2)$ can still be approximated by the scaled
chi-square distribution because the MLE is $\sqrt{n}$ consistent. To
account for this substitution, we estimate $a$ and $g$ by
replacing $\nu$ by $\tilde{\nu}$ based on the efficient
information. The elements of the Fisher information matrix of
${\blm\tau}=(\tau^{2}_{1}, \tau^{2}_{2}, \tau^{2}_{3})$ are given by
\begin{eqnarray*}
I_{{\blm\tau}{\blm\tau}}&=&\tfrac{1}{2}\left[
\matrix{
\tr(P_{0}\mathbf{K}_{1}P_{0}\mathbf{K}_{1}) & \tr(P_{0}\mathbf
{K}_{1}P_{0}\mathbf{K}_{2}) & \tr(P_{0}\mathbf{K}_{1}P_{0}\mathbf
{K}_{3})\cr
\tr(P_{0}\mathbf{K}_{2}P_{0}\mathbf{K}_{1}) & \tr(P_{0}\mathbf
{K}_{2}P_{0}\mathbf{K}_{2}) & \tr(P_{0}\mathbf{K}_{2}P_{0}\mathbf
{K}_{3})\cr
\tr(P_{0}\mathbf{K}_{3}P_{0}\mathbf{K}_{1}) & \tr(P_{0}\mathbf
{K}_{3}P_{0}\mathbf{K}_{2}) & \tr(P_{0}\mathbf{K}_{3}P_{0}\mathbf
{K}_{3})}
\right],
\\
I_{{\blm\tau}\sigma^2}&=&\tfrac{1}{2}\left[
\matrix{
\tr(P_{0}\mathbf{K}_{1}) & \tr(P_{0}\mathbf{K}_{2}) & \tr(P_{0}\mathbf
{K}_{3})}
\right]^{T}
\end{eqnarray*}
and $I_{\sigma^2\sigma^2}=\frac{1}{2}\tr(P_{0}P_{0})$. Then the efficient
information
$\tilde{I}_{{\blm\tau}{\blm\tau}}=I_{{\blm\tau}{\blm\tau}}-I_{{\blm\tau
}\sigma^2}^{T}I_{\sigma^2\sigma^2}^{-1}I_{{\blm\tau}\sigma^2}$
and
$\tilde{\nu}=\operatorname{Var}[S(\hat{\sigma}^2)]\approx \operatorname{SUM}[\tilde{I}_{{\blm\tau
}{\blm\tau}}]$,
where operator ``SUM'' indicates the sum of every element of the matrix.

\subsubsection{\texorpdfstring{Testing for G$\times$G interaction}{Testing for G x G interaction}}
For testing the interaction effect, that is, testing $H_{0}^{2}\dvtx
\tau^{2}_{3}=0$, we also apply a score test. Let
$\Sigma=\sigma^2I+\tau^{2}_{1}\mathbf{K}_{1}+\tau^{2}_{2}\mathbf{K}_{2}$.
The score function (\ref{scorefunction}) under this null hypothesis
becomes
\begin{eqnarray*}
\frac{\partial\ell_{R}}{\partial\tau^{2}_{3}}\bigg|_{\tau
^{2}_{3}=0}&=&-\frac{1}{2}[\tr(P_{01}\mathbf{K}_{3})-({\mathbf y}-\hat{\mu
}\mathbf{1})^{T}\Sigma^{-1}\mathbf{K}_{3}\Sigma^{-1}({\mathbf y}-\hat
{\mu
}\mathbf{1})]\\
&=&-\frac{1}{2}\bigl(\tr(P_{01}\mathbf{K}_{3})-{\mathbf y}^{T}P_{01}\mathbf
{K}_{3}P_{01}{\mathbf y}\bigr),
\end{eqnarray*}
where
$P_{01}=\Sigma^{-1}-\Sigma^{-1}\mathbf{1}(\mathbf{1}^{T}\Sigma
^{-1}\mathbf{1})^{-1}\mathbf{1}^{T}\Sigma^{-1}$
is the projection matrix under the null. Then
\[
S_{I}=\tfrac{1}{2}{\mathbf y}^{T}P_{01}\mathbf{K}_{3}P_{01}{\mathbf y}.
\]

Similarly, the Satterthwaite method is used to approximate the
distribution of $S_{I}$ by $a_{I}\chi^2_{g_{I}}$. Parameters $a_{I}$
and $g_{I}$ are estimated by MOM. Specifically,
$\hat{a}_{I}=\nu_{I}/2\delta_{I}$ and
$\hat{g}_{I}=2\delta_{I}^2/\nu_{I}$, where
$\delta_{I}=\frac{1}{2}\tr(P_{01}\mathbf{K}_{3})$ and
$\nu_{I}=\frac{1}{2}\tr(P_{01}\mathbf{K}_{3}P_{01}\mathbf{K}_{3})-\frac
{1}{2}\Phi^T\Delta^{-1}\Phi$, in which
\[
\Phi=\left[\matrix{\tr(P_{01}^2\mathbf{K}_{3})&
\tr(P_{01}\mathbf{K}_{3}P_{01}\mathbf{K}_{1})&
\tr(P_{01}\mathbf{K}_{3}P_{01}\mathbf{K}_{2})}\right]^T
\]
and
\[
\Delta=\left[
\matrix{
\tr(P_{01}^2) & \tr(P_{01}^2\mathbf{K}_{1}) & \tr(P_{01}^2\mathbf
{K}_{2})\cr
\tr(P_{01}^2\mathbf{K}_{1}) & \tr(P_{01}\mathbf{K}_{1}P_{01}\mathbf
{K}_{1}) & \tr(P_{01}\mathbf{K}_{1}P_{01}\mathbf{K}_{2})\cr
\tr(P_{01}^2\mathbf{K}_{2}) & \tr(P_{01}\mathbf{K}_{2}P_{01}\mathbf
{K}_{1}) & \tr(P_{01}\mathbf{K}_{2}P_{01}\mathbf{K}_{2})}
\right].
\]

\section{Simulation study}\label{simulation}
\subsection{Simulation design}
Monte Carlo simulations were conducted to evaluate the performance of
the proposed method for detecting overall genetic effects as well as
interaction between two genes.\vadjust{\goodbreak} Genotype data were simulated using the
MS program developed by \citet{Hud02}. The MS program generates
haplotype samples by using the standard coalescent approach in which
the random genealogy of a sample is first generated and the mutations
are randomly placed on the genealogy. We first simulated two
independent samples of haplotypes. Parameters of the coalescent model
were set as follows: (1) the diploid population size $N_{0}=10\mbox{,}000$;
(2) the mutation parameter $\theta=4N_{0}\mu=5.610\times10^{-4}/bp$;
and (3) the cross-over rate parameters were
$\rho=4N_{0}r=4.0\times10^{-3}/bp$ and $\rho=8\times10^{-3}/bp$ for the
two samples. In each sample, 100 haplotypes were simulated for a locus
with 10~kb long and the number of SNP sequences was set to be 100. Two
haplotypes were then randomly drawn within each simulated haplotype
pool and paired to form the genotype on the locus for an individual.
For each individual, we randomly selected 10 adjacent SNPs with minor
allele frequency (MAF) greater than $5\%$ to form a gene. This was done
separately for each simulated haplotype pool. Finally, we had genotypes
for $n$ individuals for two separate genes with 10 SNPs each, and the
two genes were independent.

%
\begin{figure}[b]

\includegraphics{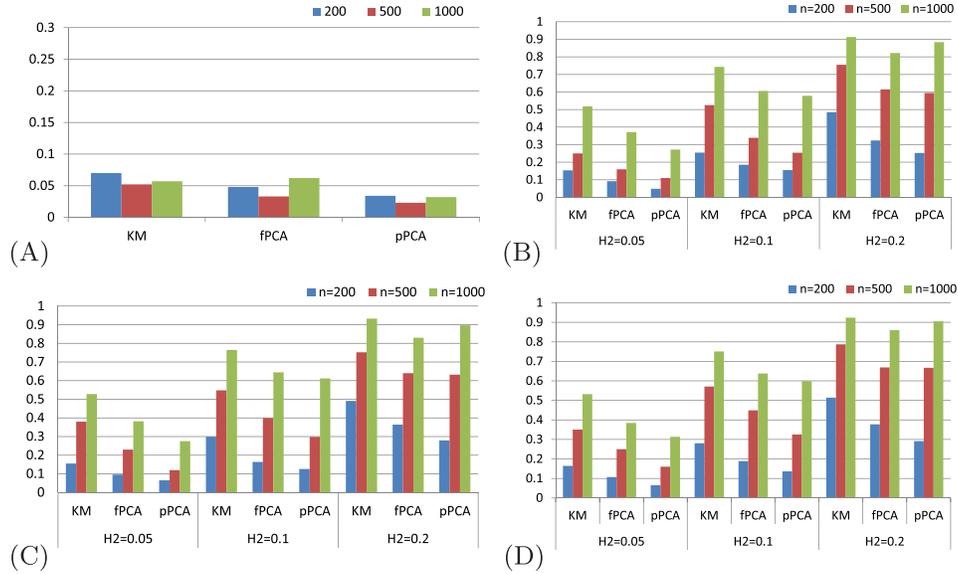}

\caption{The empirical type I error \textup{(A)} and
power \textup{(B)--(D)} of the three methods, where KM, fPCA and pPCA
refer to the proposed kernel machine method, the full and partial PCA
methods, respectively. Different heritability levels \textup{(H2)} were
assumed. The variance terms $(\sigma^{2}, \tau_{1}^{2}, \tau_{2}^{2},
\tau_{3}^{2})$ corresponding to different heritability levels
($\mbox{\textup{H2}}=0.05$, 0.1, 0.2) were given as in: \textup{(B)} (0.8,
0.021, 0.021, 0), (0.8, 0.044, 0.044, 0), (0.8, 0.1, 0.1, 0); \textup{(C)} (0.8,
0.016, 0.016, 0.088), (0.8, 0.036, 0.036, 0.018), (0.8, 0.08, 0.08,
0.04); and \textup{(D)} (0.8, 0.011, 0.011, 0.022), (0.8, 0.022, 0.022, 0.044),
(0.8, 0.05, 0.05, 0.1).}\label{power}
\end{figure}

When simulating phenotypes, four scenarios were
considered (Figure~\ref{power}). In
scenario I [Figure~\ref{power}(A)], the genetic effects were all set to
zero so that we could assess the false positive control. In
scenario II [Figure~\ref{power}(B)], we considered the main effects for
the two genes, but
set the interaction effect as zero. In scenarios~III [Figure \ref
{power}(C)] and IV [Figure~\ref{power}(D)], both
main effects and interaction effect were considered. The difference
between scenarios III and IV is that the interaction effect in
scenario III is smaller than the main effect, while in scenario IV
it is larger than the main effects. Quantitative traits of interest
were simulated from a multivariate normal distribution with mean
$\mu\mathbf{1}_{n\times1}$ and variance--covariance matrix
$V=\sigma^2\mathbf{I}+\tau^{2}_{1}\mathbf{K}_{1}+\tau^{2}_{2}\mathbf
{K}_{2}+\tau^{2}_{3}\mathbf{K}_{3}$,
where $\tau^{2}_{1}, \tau^{2}_{2}, \tau^{2}_{3}$ took different
values under different scenarios; $\mathbf{K}_{\ell}, \ell=1, 2, 3$, are
the kernel matrices using the allele matching method described
before. Different sample sizes ($n=200$, 500 and 1000) and different
heritability ($H^2=0.05$, 0.1, 0.2) were assumed. Let
$\sigma^2_G=\tau^2_1+\tau^2_2+\tau^2_3$. The heritability was
defined as $H^2=\sigma^2_G/(\sigma^2_G+\sigma^2)$. In all simulation
scenarios, we fixed the residual variance $\sigma^2=0.8$, and $\tau
_1^2$ and $\tau_2^2$ were set to be equal.

\subsection{Model comparison}
We compared our simulation results with two other methods
described in the following. \citet{Wanetal09} proposed an interaction
method using a partial least squares approach which was developed
specifically for binary disease traits. The method cannot be applied
for quantitative traits. However, in Wang et~al.'s paper they
compared their method with a regression-based principle component
analysis method. Specifically, assuming an additive model for each
marker in which genotypes AA, Aa and aa are coded as 2, 1, 0,
respectively, the singular value decomposition (SVD) can be applied
to both gene matrices. Let $G_j=(S_{1}, S_{2}, \ldots, S_{L_{j}})$ be an
$n\times L_j$ SNP matrix for
gene $j$ (\mbox{$=$}$1,2$). The SVD for $G_j$ can be expressed as
$G_j=U_jD_jV_j^T$, where $D_j$ is a diagonal\vspace*{1pt} matrix of singular
values, and the elements of the column vector $U_j$ are the
principal components $U_j^1, U_j^2,\ldots, U_j^{m_j}$ ($m_j\leq
L_j$ is the rank for $G_j$). An interaction model can be expressed
as
%
%
\begin{equation}\label{pPCA}
{\mathbf y}=\mu\mathbf{1}+\sum^{L_1}_{l_1=1}\beta_{l_1}S_{l_1}+\sum
^{L_2}_{l_2=1}\beta_{l_2}S_{l_2}+\gamma U_1^1U_1^2,
\end{equation}
where $\gamma$ represents the interaction effect between the first pair
of PCs corresponding to the largest eigenvalues in the two genes. The
main effect of each gene is modeled through the sum of all single
marker effects. For simplicity, only one interaction effect between the
first PC corresponding to the largest eigenvalues in each gene was
considered in \citet{Wanetal09}. We followed \citet
{Wanetal09} and
compared the performance of our model with this one.

In principle, one can select PCs for each gene based on the
proportion of variation explained (say, $>$85\%). Then, pairwise
interactions can be considered for all selected PCs in model
(\ref{pPCA}). Thus, if we replace the main effect of each gene in
model (\ref{pPCA}) with PCs rather than single SNPs to reduce the\vadjust{\goodbreak}
model degrees of freedom, model (\ref{pPCA}) then becomes
%
%
\begin{equation}\label{fPCA}
{\mathbf y}=\mu\mathbf{1}+\sum^{P_1}_{p_1=1}\beta_{p_1}U_{p_1}^{1}+\sum
^{P_2}_{p_2=1}\beta_{p_2}U_{p_2}^{2}+\sum_{p_1=1}^{P_1}\sum
_{p_2=1}^{P_2}\gamma_{p_1p_2} U_{p_1}^1U_{p_2}^2,
\end{equation}
where $U_{p_j}^{j}, j=1,2$, represents the PCs for gene $j$, and $P_j,
j=1,2$, is chosen based on the proportion of variation explained by
the number of PCs in gene $j$. With this regression model, we
considered all possible pairwise interactions of the selected PCs.
G$\times$G interaction was assessed by testing
$H_0\dvtx\gamma_{p_1p_2}=0$, for all $p_1$ and $p_2$. This model was
applied by \citet{HeWanEdm} in their gene-based interaction
analysis.

In addition to the two models above, we also compared our gene-centric
approach to a simple pairwise SNP interaction model. Details of the
comparison are given in Section~\ref{sresult}. For a given simulation
scenario, 1000 simulation runs were conducted. Type I error rates and
power were examined at the nominal level $\alpha=0.05$.\vspace*{-2pt}

\subsection{Simulation results}\label{sresult}\vspace*{-2pt}
\subsubsection{Comparison of the three methods for the overall genetic test}
We first evaluated the type I error rate and the power of the three
methods for testing the overall genetic effects (i.e., $H_0\dvtx\tau
_1^2=\tau_2^2=\tau_3^2=0$).
Figure~\ref{power} summarizes the comparison results between our
kernel machine (KM) method and the partial PCA (pPCA) [model~(\ref
{pPCA})] and the full PCA (fPCA) [model~(\ref{fPCA})] methods.
In Figure~\ref{power}(A), we can see that our method has
the empirical type I error rate reasonably controlled for the
overall genetic effect test. The partial PCA-based
interaction model generates very conservative results.

In all simulations we fixed the residual variance $\sigma^2$ to 0.8,
and changed the three genetic effects to get different heritability
levels. For scenario II [Figure~\ref{power}(B)], data were simulated
assuming no interaction (i.e., $\tau_3^2=0$). Then $\tau_1^2$ and $\tau
_2^2$ were calculated for a given heritability level. For example, when
$H^2=0.05$, $\tau_1^2=\tau_2^2=0.021$. In scenarios III and IV, the
interaction effect was set as either half of the main effects (scenario
III) or twice the main effects (scenario IV). As we expected, the
testing power increases as the heritability level and sample size
increased [Figure~\ref{power}(B)--(D)]. For a fixed heritability level, the
KM method always outperformed the other two under different sample
sizes. The power difference of the KM method over the other two is more
striking under a small heritability level (say, 0.05) or when the
sample size is small.
For most complex diseases in humans, the heritability of a genetic
variant is generally small. Thus, the KM method is preferable over the
other two in the first stage of the interaction analysis. In addition,
we also observed that the KM method is insensitive to whether the
genetic effect is due to the main effect or the interaction, whereas
the PCA-based method gains power as more of the genetic effect is due
to the interaction.\vadjust{\goodbreak}

\subsubsection{Comparison of the three methods for the interaction test}
Interaction may be due to a variety of underlying mechanisms. Some
genes might have both significant main effects and interaction
effect, while others might only incur interaction effect without
main effects. Simulation studies were designed to compare the
performance of the proposed KM method in discovering
gene${}\times{}$gene interactions over the other two methods, considering
different interaction
effect sizes. Since the power of an interaction test is largely
determined by the size of the interaction effect, we simulated data
assuming different proportions of interaction effects among the total
genetic variance. This proportion is defined by
$\eta=\tau_{3}^2/\tau^2$, where $\tau^2$ $(\mbox{$=$}\tau_{1}^2+\tau_{2}^2+\tau
_{3}^2)$ refers to the total genetic variance. For a fixed total
genetic variance, the value of $\eta$ indicates the strength of the
interaction effect between two genes. For a fixed residual variance
($\sigma^2=0.8$), the total genetic variance is set to 0.2 when
$H^2=0.2$ and to 0.53 when $H^2=0.4$. The variance size for the two
main effects were set to be equal, so we could calculate the
interaction variance. For
example, $(\tau_{1}^2, \tau_{2}^2, \tau_{3}^2)=(0.08, 0.08, 0.04)$ when
$\eta=0.2$, and $(\tau_{1}^2, \tau_{2}^2,
\tau_{3}^2)=(0.02, 0.02, 0.16)$ when $\eta=0.8$ under $H^2=0.2$. Six
values of the
proportion $\eta=(0, 0.2, 0.4, 0.6, 0.8, 1.0)$ were considered,
including the two extreme cases: no interaction at all $(\eta=0)$ and
pure interaction $(\eta=1)$. The method was compared with
the other two PCA-based interaction analyses under two different sample
sizes, 500 and 1000.

%
%
\begin{figure}[b]

\includegraphics{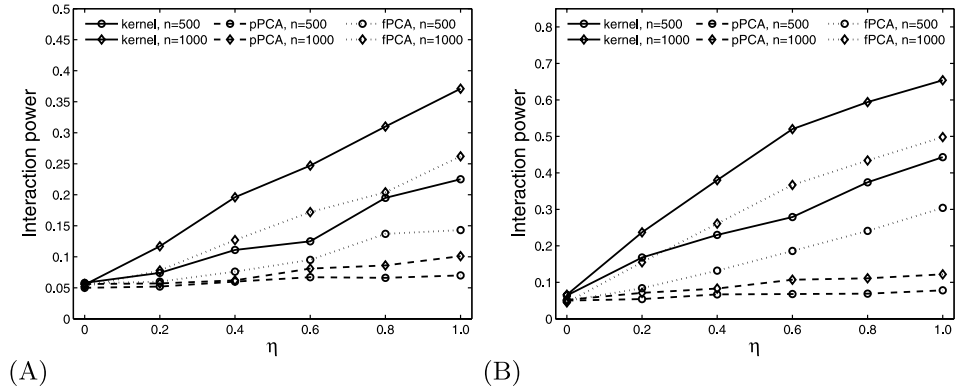}

\caption{Power comparison of the proposed KM model
(solid line), pPCA model (\protect\ref{pPCA}) (dashed line) and fPCA
model (\protect\ref{fPCA}) (dotted line) under different sample sizes
($n$) and different interaction sizes ($\eta$), where $\eta$ refers to
the proportion of interaction effect among the total genetic effect.
\textup{(A)} $H^2=0.2$ and \textup{(B)} $H^2=0.4$.}\label{ipower}
\end{figure}

Figure~\ref{ipower} shows power comparison (based on 1000 replicates)
under two different heritability levels [Figure~\ref{ipower}(A) for
$H^2=0.2$ and Figure~\ref{ipower}(B) for $H^2=0.4$].
The type I error (when $\eta=0$) for the interaction test is reasonably
controlled for the three methods. As we expected, the
interaction power increases as\vadjust{\goodbreak} the interaction effect size ($\eta$)
increases. Among the three methods, our KM
method has the highest power. The partial PCA model (\ref{pPCA}) has
the lowest
power. This demonstrates that only considering interaction of the first
principle component in each gene is not enough to capture the
interaction between two genes. The effect of sample size on the
interaction power is also significant. Large sample size always leads
to large power.

A common concern in detecting gene--gene interactions is the
computational burden. As we focus on genes as testing units, the total
number of tests reduces dramatically in a gene-centric analysis
compared to a single-marker based analysis. In addition, computation
time can be saved by implementing the two-stage analysis: (1) doing
score tests in the first stage to assess the significance of all
genetic components; and (2) testing interaction only for those pairs
showing statistical significance after multiple testing adjustment in
the first stage. The computation for the first step is very fast since
only parameters under the null model, which is a regular linear
regression model, need to be estimated. In general, only a small
fraction of signals can pass the significant threshold in the first
stage, leaving the second stage interaction test with less
computational burden. For a quick comparison of the KM method with the
PC-based approaches, when $n=500$, the computation time with 1000
simulation replicates for KM, pPCA and fPCA are 28 min, 24.5 min and
24 min, respectively, for the overall test. When testing the interaction
term, the KM method takes about 9sec for a single run on average, and
is slower than the other two methods. However, this should not be a
concern in real applications since the number of interactions is
generally not very large after the first stage screening.

In summary, our KM method outperforms the other two methods in the
overall genetic test as well as in the interaction test under different
simulation scenarios. The results also indicate that large sample sizes
are needed for the detection of the interaction term compared to the
detection of main effects. Even though the PCA-based analysis has been
applied for candidate gene-based association analyses
[\citet{WanAbb08}, \citet{Wanetal09}], our simulation studies
show that it may
not be suitable for interaction analysis.

\subsubsection{Comparison with a single SNP interaction model}
In a regression-based analysis for interaction, the commonly used
approach is the single SNP interaction model with the form
%
%
\begin{equation}\label{singleSNP}
y_{i}=\beta_0+\beta_1S_{1i}+\beta_2S_{2i}+\beta
_{12}S_{1i}S_{2i}+\varepsilon_{i},\qquad i=1, 2,\ldots, n,
\end{equation}
where $\beta_0$ is the intercept; $\beta_1$, $\beta_2$ and $\beta_{12}$
represent the effects of SNP $S_1$ in gene~1, SNP $S_2$ in gene 2 and
the interaction effect between the two, respectively; $\varepsilon
_{i}\sim N(0,\sigma^2)$. We simulated data according to model (\ref
{singleSNP}) assuming a MAF $p_A=0.3$. Different heritabilities and
different sample sizes were assumed. For simplicity, we assumed the
same effect size for the three coefficients which are calculated under\vadjust{\goodbreak}
specific heritability ($H^2=0.2$ and~0.4) when generating the data. We
considered an extreme case in which each gene only contains one single
SNP. Data generated with model (\ref{singleSNP}) are subject to both
the single SNP interaction and the proposed kernel interaction
analysis. With this simulation we tried to assess how robust the KM
method is when there is only one pair of functional SNPs in two genes.

%
%
\begin{table}
\caption{List of empirical type I error and power based on 1000
simulation runs when~data were~simulated with model
(\protect\ref{singleSNP})}\label{single}
\begin{tabular*}{\tablewidth}{@{\extracolsep{\fill}}lcc
c@{\hspace*{45pt}}ccc@{\hspace*{-16pt}}}
\hline
& \multicolumn{1}{c}{\multirow{2}{67pt}[-8pt]{\centering{\textbf{Coefficients} $\bolds{(\beta_0, \beta_1, \beta_2, \beta
_{12})}$}}} &
\multicolumn{1}{c}{\multirow{2}{50pt}[-8pt]{\centering{\textbf{Sample size} $\bolds{(n)}$}}}
& \multicolumn{2}{c}{\textbf{Single SNP
interaction}}&\multicolumn{2}{c@{}}{\textbf{Kernel interaction}}\\[-4pt]
&&& \multicolumn{2}{c}{\hrulefill} & \multicolumn{2}{c@{}}{\hrulefill}\\
$\bolds{H^2}$& &&\multicolumn{1}{c}{$\bolds{P_{o}}$\hspace*{35pt}}
&\multicolumn{1}{c}{$\bolds{P_{i}}$\hspace*{5pt}}
&\multicolumn{1}{c}{$\bolds{P_{o}}$}&\multicolumn{1}{c@{\hspace*{-16pt}}}{$\bolds{P_{i}}$}\\
\hline
0.2&(0.19, 0, 0, 0)&\hphantom{0}200&0.055&0.019&0.059&0.003\\
&&\hphantom{0}500&0.058&0.019&0.057&0.003\\
&&1000&0.052&0.017&0.059&0.003\\
[3pt]
&(0.19, 0.19, 0.19, 0)&\hphantom{0}200&0.497&0.03\hphantom{0}&0.534&0.032\\
&&\hphantom{0}500&0.923&0.045&0.911&0.046\\
&&1000&0.999&0.048&0.997&0.053\\
[3pt]
&(0.19, 0.19, 0.19, 0.19)&\hphantom{0}200&1\hphantom{.000}&0.221&1\hphantom{.000}&0.183\\
&&\hphantom{0}500&1\hphantom{.000}&0.419&1\hphantom{.000}&0.349\\
&&1000&1\hphantom{.000}&0.714&1\hphantom{.000}&0.635\\
[6pt]
0.4&(0.51, 0, 0, 0)&\hphantom{0}200&0.053&0.022&0.053&0.003\\
&&\hphantom{0}500&0.049&0.016&0.062&0.001\\
&&1000&0.054&0.024&0.057&0.008\\
[3pt]
&(0.51, 0.51, 0.51, 0)&\hphantom{0}200&1\hphantom{.000}&0.051&1\hphantom{.000}&0.058\\
&&\hphantom{0}500&1\hphantom{.000}&0.062&1\hphantom{.000}&0.067\\
&&1000&1\hphantom{.000}&0.054&1\hphantom{.000}&0.058\\
[3pt]
&(0.51, 0.51, 0.51, 0.51)&\hphantom{0}200&1\hphantom{.000}&0.850&1\hphantom{.000}&0.648\\
&&\hphantom{0}500&1\hphantom{.000}&0.996&1\hphantom{.000}&0.964\\
&&1000&1\hphantom{.000}&1\hphantom{.000}&1\hphantom{.000}&1\hphantom{.000}\\
\hline
\end{tabular*}
\legend{$P_o$ and $P_i$ refer to the power for testing the overall
genetic effects (i.e., $H_0\dvtx\tau_1^2=\tau_2^2=\tau_3^2=0$ for the
kernel approach and $H_0\dvtx\beta_1=\beta_2=\beta_{12}=0$ for the
pairwise\vspace*{1pt} SNP interaction analysis) and for testing
interaction effect (i.e., $H_0\dvtx\tau_3^2=0$ for the kernel approach
and $H_0\dvtx\beta _{12}=0$ for the pairwise SNP interaction analysis),
respectively.}\vspace*{-3pt}
\end{table}

Table~\ref{single} shows that both models show comparable type I error
control for the overall genetic test (see $P_o$ in the table). For the
interaction test, it looks like the kernel approach generates more
conservative results. Here the interaction test is nested within the
overall genetic test. If we aggregate the results by dividing $P_i$
by~$P_o$, the single SNP analysis actually produces more inflated false
positives compared to our kernel approach when no genetic effect is
involved. When data were simulated assuming only main effects but no
interaction (case $\beta_{12}=0$), the two approaches yield very
similar false positive rates, indicating reasonable performance of the
kernel approach for false positive control.\vadjust{\goodbreak}

For the power analysis, we found little difference between the two
methods for the overall genetic test ($P_o$), especially under large
sample size and high heritability level. For the interaction test
($P_i$), the power increases as sample size and heritability level
increase. We observed relatively large power differences between the
models when the sample size is small ($<$500). As sample size
increases, the difference diminishes. For example, the power difference
is 0.2 under $n=200$ and $H^2=0.4$, which is reduced to 0.032 when $n$
increases to 500 and to zero when $n=1000$. When more than one
functional SNP within each gene is involved in interacting with one
another to affect a trait variation, the kernel method consistently
outperformed the single SNP interaction model (data not shown). In
summary, our model performs reasonably well in different scenarios
compared to the other methods. Even when there is only one single SNP
pair interacting with each other in two genes, our analysis produces
results as good as the ones analyzed with the true model, especially
under large sample sizes and high heritability (Table~\ref{single}).

\section{Applications to real data}\label{case}
\subsection{Application \textup{I}: Analysis of birth weight data}
A candidate gene study was initially conducted in order to
study genetic effects associated with being large for
gestational age (LGA) and small for gestational age (SGA). Subjects
were recruited through the Department of Obstetrics and Gynecology
at Sotero del Rio Hospital in Puente Alto, Chile, and SNPs were
selected for genotyping in order to capture at least $90\%$ of the
haplotypic diversity of each gene. Individuals were genotyped
at 797 SNP markers in 186 unique candidate genes. Missing genotypes
were imputed using a conditional probability approach as described in
\citet{Cuietal08}. We combined the two data sets
(LGA and SGA) and used baby's birth weight (in kg) as the response
variable to assess if there are any genes or interaction of genes
that could explain the normal variation of new born baby's birth
weight. The case/control classification in the two data sets was based
on baby's birth weight together with mother's gestational age. A total
number of 1511 individuals in the case and control combined data set
were used for analysis after removing 14 individuals with birth weight
3${}\times{}$IQR (inter-quartile
range) above $Q_3$ or below $Q_1$. After removing these extreme
observations and a Box--Cox transformation, the distribution of the
birth weight data in the combined sample was approximately normal.

%
\begin{table}
\caption{List of gene pairs with $p$-value less than 0.001 in the overall
genetic effect test. Gene pairs with significant interaction effect
($p$-value$_i<$0.05) are indicated with bold font}\label{BW}
\begin{tabular*}{\tablewidth}{@{\extracolsep{\fill}}lcllcccl@{}}
\hline
\textbf{Gene 1} & \textbf{Gene 2} &\multicolumn{1}{c}{$\bolds{\tau_1^2}$}
&\multicolumn{1}{c}{$\bolds{\tau_2^2}$}&\multicolumn{1}{c}{$\bolds{\tau_3^2}$}
&\multicolumn{1}{c}{$\bolds{\sigma^{2}}$} &
\multicolumn{1}{c}{$\bolds{p}$\textbf{-value}}
& \multicolumn{1}{c@{}}{$\bolds{p}$\textbf{-value}$\bolds{_i}$}\\
\hline
{\bf ANG} & {\bf EDN1} & 0.0340 &9.1E--08 & 0.0024 & 0.3199 & 0.000656 &
8.07E--06\\\\
\textbf{PDGFC} & COL5A2 & 0.0095 & 2.6E--06 & 0.0051 & 0.3232 & 0.000494
& 0.2410\\
& F3 & 0.0056 & 1.5E--06 & 0.0061 &0.3242 & 0.000781 & 0.0506\\
& GP1BA & 0.0118 & 0.0362 &$\mbox{$<$}10^{-6}$ & 0.3229 & 0.000283 & 0.7462\\
& IGF1& 0.0124 & 0.0090 & $\mbox{$<$}10^{-6}$ & 0.3234 & 0.000259 & 0.5133\\
& IL1B & 0.0118 & 0.0049 & $\mbox{$<$}10^{-6}$ & 0.3227 & 0.000554 & 0.6228\\
& IL9&1.0E--07 & 0.0151 & 0.0129 & 0.3210 &0.000066 & 0.3849\\
& LPA &0.0130 &0.0076 & $\mbox{$<$}10^{-6}$ & 0.3226 & 0.000294 & 0.5231\\
& MMP7& 0.0122 & 0.0051 & $\mbox{$<$}10^{-6}$ & 0.3236 & 0.000518 & 0.6548\\
& OXTR & 0.0007 & 1.1E--06 & 0.0124 & 0.3218 & 0.000447 & 0.2107\\
& PLAUR & 0.0128 & 0.0396 & $\mbox{$<$}10^{-6}$ & 0.3194 &0.000536 & 0.5460\\
& {\bf PTGER3}& 0.0057 & 1.7E--07 & 0.0051 & 0.3240 & 0.000279 & 0.0434\\
& PTGS2 & 0.0127 & 0.0058 & $\mbox{$<$}10^{-6}$ & 0.3226 & 0.000514 & 0.7092\\
& TIMP2 & 0.0075 & 1.1E--07 & 0.0043 & 0.3238 & 0.000916 & 0.2351\\
& TLR4 & 0.0122 & 0.0115 & $\mbox{$<$}10^{-6}$ & 0.3239 & 0.000581 & 0.5606\\ \\
\textbf{PTGS2} & ANG& 0.0062 & 0.0182 & $\mbox{$<$}10^{-6}$ & 0.3218 & 0.000416
& 0.7016\\
& EDN1& 0.0054 & 0.0030 & $\mbox{$<$}10^{-6}$ & 0.3243 & 0.000730 & 0.7966 \\
& LPA &0.0055 & 0.0062 & $\mbox{$<$}10^{-6}$ & 0.3239 & 0.000988 & 0.5328\\
& PDGFB & 0.0010 &1.5E--06 & 0.0042 & 0.3246 &0.000782 & 0.2739 \\
& \textbf{PGF} & 0.0261 &7.4E--08 & 0.0044 & 0.3240 & 0.000850 &0.0062\\
& \textbf{PLAU} & 0.0004 & 3.0E--06 & 0.0057 & 0.3231 & 0.000260 &
0.0207\\\\
\textbf{IL9} &GP1BA & 1.1E--06 & 0.0082 & 0.0274 & 0.3220 & 0.000936 &
0.4592\\
&\textbf{IGF1}& 0.0174 & 2.8E--08 & 0.0075& 0.3227 & 0.000540 & 0.0009\\
\hline
\end{tabular*}
\legend{$\sigma^2$ is the residual variance; $p$-value is for testing $H_0\dvtx
\tau_1^2=\tau_2^2=\tau_3^2=0$ and $p$-value$_i$ is for testing $H_0\dvtx
\tau
_3^2=0$. Note that the estimated interaction effect size does not
necessarily indicate how strong an interaction is. The strength of an
interaction is rather reflected by the interaction testing $p$-value
(denoted by $p$-value$_i$).}
\end{table}

A two-dimensional pairwise G${}\times{}$G interaction search was
conducted (total 17,205 gene pairs). The score test for testing
$H^1_0\dvtx\tau^2_1=\tau^2_2=\tau^2_3=0$ was determined and $p$-values were
obtained for all gene pairs.
For a two-dimensional search, it is not clear how to set up a genome-wide
threshold to correct for multiple testings. Obviously, the 17,205 tests
are not independent and it is too stringent to use the Bonferroni
correction method. Thus, we used an arbitrary threshold of 0.001 as a
cutoff. Totally, 23 gene pairs were\vadjust{\goodbreak} found to be significant
with this cutoff. A detailed list of these gene pairs, their effect
estimates and the $p$-values for the overall genetic and interaction
test are shown in Table~\ref{BW}. Among the 23 gene pairs, five
significant G${}\times{}$G interactions were detected at the 0.05
level. These are gene pairs ANG--EDN1, PDGFC--PTGER3, PTGS2--PGF,
PTGS2--PLAU and IL9--IGF1 (indicated by bold fonts in Table~\ref{BW}). If
we increase the overall testing $p$-value threshold (to 0.0005), only 9
gene pairs will remain in the list and only 2~gene pairs will show significance.

The results indicate a strong genetic effect for gene PDGFC
(Platelet-derived growth factor C). This gene is a key component of the
PDGFR-$\alpha$ signaling pathway. Studies have shown that PDGFC
contributes to normal development of the heart, ear, central nervous
system (CNS) and kidney [\citet{ReiVarLil05}]. Even though its main
effect is very strong, the only gene found to have significant
interaction effect with this gene was gene PTGER3 ($p$-value$_i=0.0434$).

Among the five interacting gene pairs, the interaction between genes
ANG (Angiogenin) and EDN1 (Endothelin 1) shows the strongest
interaction signal ($p$-value$_i<10^{-5}$). Study has shown that
dysregulation of angiopoietins is associated with low birth weight
[\citet{Siletal10}]. \citet{Nezetal09} studied the role of
endothelin 1 in preeclampsia and nonpreeclampsia women, and found
that EDN1 correlates with the degree of fetal growth restriction.
Although no study has reported the interaction between the two
genes, our finding suggests a potential role of interaction between
the two genes in affecting fetal growth. Further functional analysis
is needed to validate this result.

Interactions were also found between gene PTGS2
(Prostaglandin-endo\-peroxide synthase 2) and genes PLAU (Urokinase-type
plasminogen activator) and PGF (Placental growth factor), and between
gene IL9 (Interleukin~9) and IGF1 (Insulin-like growth factor 1). It
has been recognized that genes PGF and IGF1 are associated with fetal
growth [\citet{Toretal03}, \citet{Osoetal96}]. The identification
of interactions of the two genes with other genes provides important
biological hypotheses for further lab verification.

\subsection{Application \textup{II}: Analysis of yeast eQTL data}
The second data set we analyzed with our model is a well studied
yeast eQTL mapping data set generated to understand the genetic
architecture of gene expression [\citet{BreKru05}]. The data
were generated from 112 meiotic recombinant progenies of two yeast
strains: BY4716 (BY: a laboratory strain) and RM11-1a (RM: a natural
isolate). The data set contains 6229 gene expression traits and 2956
SNP marker genotype profiles. As an example to show the utility of
our approach to an eQTL mapping study, we picked the expression
profile of one gene (BAT2) as the quantitative response to identify
potential genes or gene--gene interactions that regulate the
expression of this gene. Note that one of the parental
strains, RM11-1a, is a LEU2 gene knockout strain. Thus, we expect strong
segregation of this gene in the mapping population. Since BAT2 is in
the downstream of the Leucine Biosynthesis Pathway (LEU2 is in the
upstream of the same pathway), this motivates us to pick BAT2 as the
response [see Figure 5(a) in \citet{SunYuaLi08}]. The Bonferroni
correction was applied to
adjust multiple testings for the 1,072,380 gene pairs. An overall test
for pairs of gene effects was conducted followed by the score test
for interaction if the overall test is significant.

%
\begin{figure}

\includegraphics{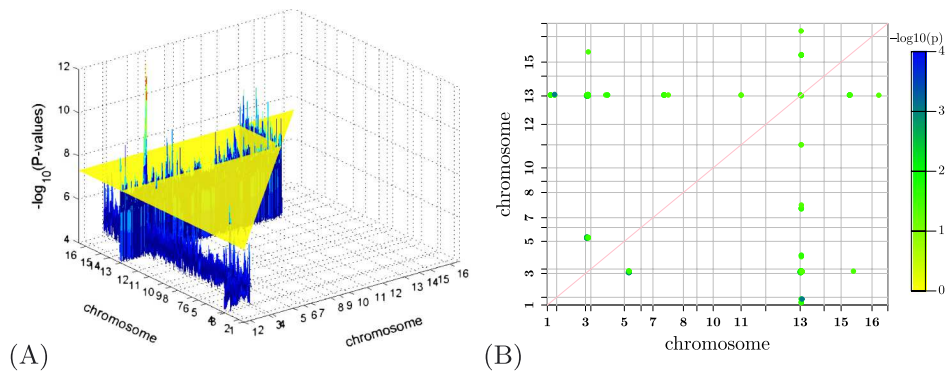}

\caption{Analysis of the expression profile for BAT2. The $-{\log10}$
transformed $p$-value profile plot of all gene
pairs for the overall test \textup{(A)} and the interaction test \textup{(B)}. The yellow
hyperplane in \textup{(A)} represents the Bonferroni
cutoff.}\label{feQTL}
\end{figure}

There were 1465 genes with some containing a single SNP marker. All the
genes were subject to the proposed kernel interaction analysis.\vadjust{\goodbreak} Figure~\ref{feQTL}(A)
shows the pairwise interaction plot for $-\log10$ transformed
$p$-values associated with the overall genetic test. The yellow
hyperplane indicates the Bonferroni correction threshold. Data points
with $p$-values larger than 10$^{-4}$ were masked. The plot indicates a
strong genetic effect at chromosomes~3 and~13, which implies that the
two locations are potential regulation hotspots. In checking the recent
literature, we found that the two positions were reported as eQTL
hotspots in a number of studies [e.g., \citet{Breetal02},
\citet{Peretal07}, \citet{LiLuCui10}].

\begin{figure}

\includegraphics{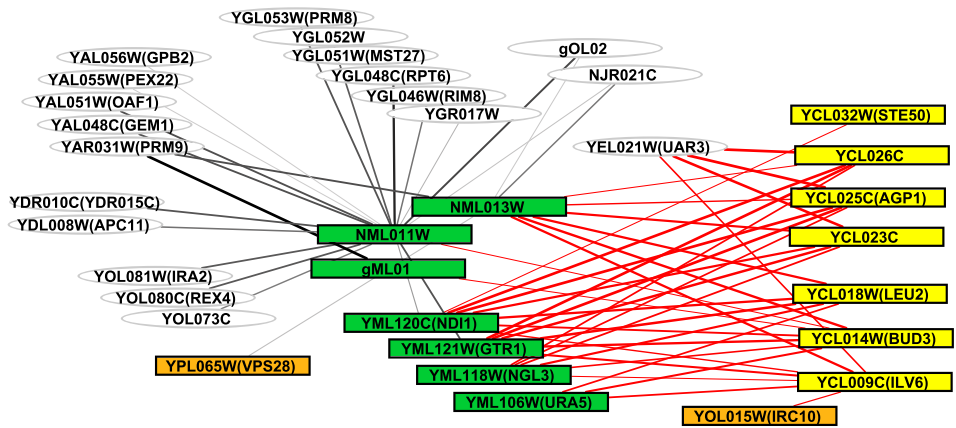}

\caption{The network graph of interacting genes for the eQTL
analysis. Each node represents a gene; nodes with light oval shape
indicate no main effect. Interacting genes are connected with lines
(the thicker the lines, the stronger the interaction). Genes with
different colors are located in different chromosomes with significant
main effects.}\label{network}\vspace*{-3pt}
\end{figure}

Out of the 1072380 gene pairs, 87 pairs were found to have
significant interactions at an experiment-wise significant level of
0.05. Figure~\ref{feQTL}(B) plots the pairwise significant
interactions. Circles correspond to significant interaction pairs,
with the darkness of the color indicating the strength of the
interaction. We saw a strong interaction pattern on chromosome 13.
One or several genes at this location interact with many other genes
to affect the transcription of gene BAT2. Another interaction
``hotspot'' is at chromosome~3 where genes (containing LUE2 and its
neighborhood genes) interact with genes at chromosomes 5, 13 and 15
to regulate BAT2 expression. We used Cytospace [\citet{ShaMarOzi03}]
to generate an interaction network map (see Figure~\ref{network}). Each
node represents a gene and the thickness of the connection line
indicates the strength of the interaction effect. Genes at the same
chromosome location are clustered together in the plot. Light nodes
with oval shapes indicate weak or no main effect. We found
strong main effects for genes on chromosomes 3 and 13. The
strongest interaction effect is between genes on chromosome 3 and
chromosome 13. We also highlighted (red lines) the interaction
between genes on chromosome 3 and others. Among the genes with no
main effects (light oval nodes), URA3 is known
as a~transcription factor [\citet{RoyExiLos90}]. Even though it\vadjust{\goodbreak}
does not
show any main effect, it shows interactions with several other genes on
chromosome
3 to regulate the expression of BAT2. The results also imply the
important role of several loci on chromosome~13. As their
functions are unknown, they could be potential candidate genes for
further lab validation.

%

\section{Discussion}\label{discussion}
The importance of gene--gene interaction in complex traits has
stimulated enormous discussion. Fundamental works in statistical
meth\-odology development have been broadly pursued in this area
[reviewed in \citet{Cor09}]. Previous investigations have
demonstrated the importance of a gene-centric approach in genetic
association studies by simultaneously considering all markers in a gene
to boost association power [e.g., \citet{Cuietal08},
\citet{BuiMar}, \citet{Lietal09}, \citet{Maetal10}].
This motivates us to develop a gene-centric approach to understand
gene--gene interactions associated with complex diseases. In this
paper, we propose a kernel machine method for a gene-centric gene--gene
interaction analysis. We adopt a~smoothing spline-ANOVA decomposition
method to decompose the genetic effects of two genes into separate main
and interaction effects, and further model and test the genetic effects
in the reproducing kernel Hilbert space. The joint variation of SNP
variants within a gene is captured by a~properly defined kernel
function, which enables one to model the interaction of two genes in a
linear reproducing Hilbert space by a cross-product of two kernel
functions. The kernel machine method is shown mathematically to be
equivalent to a linear mixed effects model. Thus, testing main and
interaction effects can be done by testing the significance of
different variance components. Extensive simulations and analysis of
two real data sets demonstrate the utility of the gene-centric
interaction analysis.\vadjust{\goodbreak}

He et~al. (\citeyear{HeWanEdm}) previously proposed a gene-based
interaction method in which each gene is summarized by several
principle components and interaction was tested through the modeling of
the PC terms rather than single SNPs. The authors proposed a weighted
genotype scoring method using pairwise LD information to test
gene--gene interaction. Their method is similar to several other
methods which jointly consider information contributed by multiple
markers [e.g., \citet{Chaetal06}, \citet{ChaCla07}]. Our
method is fundamentally different from their approach in which we
capture the joint variation of SNP variants within and between genes by
kernel functions [see \citet{Sch10N1} for more discussion of the
advantage of the kernel methods]. Our method can also be extended to
test interaction of variants by incorporating various weighting
functions to define a kernel measure [\citet{Wuetal11}], and will
be considered in our follow-up investigation. It is worth mentioning
some other methods for a gene-based analysis such as the CTGDR method
proposed by \citet{Maetal10} which accounts for the gene and
SNP-within-gene hierarchical structure. The method can identify
predictive genes as well as predictive SNPs within identified genes.
Other methods such as the grouped lasso method can also be applied for
a~gene-based analysis [\citet{MaSonHua07}]. However, their
extension to a~gene interaction analysis is not straightforward and
warrants further investigation.

The advantage of the gene-centric gene--gene interaction analysis was
previously discussed in He et~al. (\citeyear{HeWanEdm}), including reducing
the number of hypothesis tests in a genome-wide scan compared to a
single SNP interaction analysis. However, we
should not over-emphasize the role of gene-centric analysis. Our
simulation study shows that when the underlying truth is a~single SNP
pair interaction in two genes,
single-SNP interaction analysis performs better. This result agrees
with the conclusion made by He et~al. (\citeyear{HeWanEdm}). Therefore, we
recommend that investigators conduct both types of analyses (single SNP
and gene-centric) in real applications, especially when no
prior knowledge is available on how SNPs function within a gene as
well as between genes. For a large-scale genome-wide or candidate
gene study, one can also use the gene-centric approach as a
screening tool, then further target which SNPs in different genes
interact with each other.

The choice of kernel function may have potential effects on the
testing power [Schaid (\citeyear{Sch10N1}, \citeyear{Sch10N2})]. In
this paper, we consider the
AM kernel. Other kernel functions can
also be applied as discussed in detail in \citet{Sch10N2}. It is
not the
purpose of this paper to compare the
performance of difference kernel choices on the power of an
association test. A comparison study of different kernel functions
on the power of the interaction test will be considered in future
investigations.

The proposed method considers genes as testing units to test their
interaction. It is easy to extend the idea to incorporate other
genomic features such as pathways as testing units to assess\vadjust{\goodbreak}
pathway--pathway interaction under the proposed framework. The
mapping results can then be visualized by some network graphical
tools such as the Cytospace software [\citet{ShaMarOzi03}] which can
help investigators generate important biological hypotheses for
further lab validation. The computational code written in R
is available as an R-package from
\url{http://cran.r-project.org/package=SPA3G}.

%
\begin{appendix}\label{app}
\section{ANOVA decomposition}\label{Appendix1}
Suppose a function $f$ is on domain $\Gamma=\Gamma_{1}\otimes\Gamma
_{2}\otimes\Gamma_{3}$. Define
corresponding averaging operator $A_{\gamma}$ on each generic domain
$\Gamma_{\gamma}, \gamma=1, 2, 3$. An ANOVA decomposition of a
function $f$ can be obtained:
\begin{eqnarray*}
f&=&\Biggl\{\prod_{\gamma=1}^{3}(I-A_{\gamma}+A_{\gamma})\Biggr\}f\\
&=&\{
(I-A_{1})(I-A_{2})(I-A_{3})+(I-A_{1})(I-A_{2})A_{3}\\
&&\hphantom{\{}{}+(I-A_{1})A_{2}(I-A_{3})+A_{1}(I-A_{2})(I-A_{3})\\
&&\hphantom{\{}{}+(I-A_{1})A_{2}A_{3}+A_{1}A_{2}(I-A_{3})+A_{1}(I-A_{2})A_{3}+A_{1}A_{2}A_{3}\}
f.
\end{eqnarray*}
For a nested domain
$(\Gamma_{1}\otimes\Gamma_{2})\otimes\Gamma_{3}$, let $A_{12}$ be
the averaging operator on domain $(\Gamma_{1}\otimes\Gamma_{2})$.
Then the ANOVA decomposition becomes
\[
f=\{(I-A_{12})(I-A_3)+A_{12}(I-A_3)+(I-A_{12})A_3+A_{12}A_3\}f.
\]
Since
\[
(I-A_1)(I-A_2)+(I-A_1)A_2+A_1(I-A_2)=I-A_1A_2
\]
by letting $A_{12}=A_{1}A_{2}$,
\begin{eqnarray*}
&&
\{(I-A_{12})(I-A_3)+A_{12}(I-A_3)+(I-A_{12})A_3+A_{12}A_3\}f\\
&&\qquad=\Biggl\{\prod
_{\gamma=1}^{3}(I-A_{\gamma}+A_{\gamma})\Biggr\}f.
\end{eqnarray*}
Recursively, it shows that the ANOVA decomposition can also be
conducted on products of nested domains.

\section{The dual representation}\label{Equivalence}
Consider the linear mixed effect model:
\[
y=\mu\mathbf{1}+m_{1}+m_{2}+m_{12}+\epsilon,
\]
where $m_{1}, m_{2}, m_{12}$ are independent $n\times1$ vector of
random effects; $m_{1}\sim N(\mathbf{0},
\tau^{2}_{1}\mathbf{K}_{1})$, $m_{2}\sim N(\mathbf{0},
\tau^{2}_{2}\mathbf{K}_{2})$, $m_{12}\sim N(\mathbf{0},
\tau^{2}_{3}\mathbf{K}_{3})$, and $\epsilon\sim N(\mathbf{0}, \sigma
^2I)$ is
independent of $m_{1}, m_{2}$ and $m_{12}$. Henderson's normal
equation for obtaining the BLUPs of the random effects is
%
%
\begin{equation}\label{Henderson}\qquad
\left[
\matrix{
n& \mathbf{1}^{T} & \mathbf{1}^{T} & \mathbf{1}^{T} \cr
\mathbf{1}& \mathbf{I}+\dfrac{\sigma^2}{\tau_{1}^2}\mathbf{K}_{1}^{-1} &
\mathbf{I} & \mathbf{I}\vspace*{2pt}\cr
\mathbf{1}& \mathbf{I} & \mathbf{I}+\dfrac{\sigma^2}{\tau_{2}^2}\mathbf
{K}_{2}^{-1}& \mathbf{I}\vspace*{2pt}\cr
\mathbf{1}& \mathbf{I} & \mathbf{I} & \mathbf{I}+\dfrac{\sigma^2}{\tau
_{3}^2}\mathbf{K}_{3}^{-1}}
\right]
\left[
\matrix{
\mu\cr m_{1} \cr m_{2}\cr m_{12}}
\right]=\left[
\matrix{
\mathbf{1}^{T}\cr
\mathbf{I}\cr
\mathbf{I}\cr
\mathbf{I}}
\right]{\mathbf y}.
\end{equation}
It can be shown that this normal equation is equivalent to the first
order condition for estimating function $m$, equation
(\ref{sssystem}). Multiplying both sides of (\ref{sssystem})
by the matrix,
\[
\left[
\matrix{
1 & \mathbf{0} & \mathbf{0} & \mathbf{0} \cr
\mathbf{0} & \mathbf{K}_{1}^{-1} & \mathbf{0} & \mathbf{0}\cr
\mathbf{0} & \mathbf{0} & \mathbf{K}_{2}^{-1}& \mathbf{0}\cr
\mathbf{0} & \mathbf{0} & \mathbf{0} & \mathbf{K}_{3}^{-1}}
\right],
\]
one can get
\[
\left[
\matrix{
n & \mathbf{1}^{T}\mathbf{K}_{1} & \mathbf{1}^{T}\mathbf{K}_{2} &
\mathbf{1}^{T}\mathbf{K}_{3} \cr
\mathbf{1} & \mathbf{K}_{1}+\lambda_{1}\mathbf{I} & \mathbf{K}_{2} &
\mathbf{K}_{3}\cr
\mathbf{1} & \mathbf{K}_{1} & \mathbf{K}_{2}+\lambda_{2}\mathbf{I}&
\mathbf{K}_{3}\cr
\mathbf{1} & \mathbf{K}_{1} & \mathbf{K}_{2} & \mathbf{K}_{3}+\lambda
_{3}\mathbf{I}}
\right].
\left[
\matrix{
\mu\cr C_{1} \cr C_{2}\cr C_{3}
}
\right]=\left[
\matrix{
\mathbf{1}^{T}\cr
\mathbf{I}\cr
\mathbf{I}\cr
\mathbf{I}}
\right]{\mathbf y}.
\]
Letting $m_{\ell}=\mathbf{K}_{\ell}C_{\ell}$ $(\ell=1, 2)$,
$m_{12}=\mathbf{K}_{3}C_{3}$ and
$\tau_{\ell}^2=\sigma^2/\lambda_{\ell}$ $(\ell=1, 2, 3)$, the system is
exactly equation (\ref{Henderson}), which is Henderson's
normal equation of linear mixed effects model (\ref{LMM}).

\section{Calculation of the kernel matrix: An~example}\label{example}
As an example, we consider three unrelated individuals. Suppose
there are 10~SNP markers in one gene. The genotypes for these
markers are numerically coded as 0, 1 or 2 depending on the copy number
of a certain allele (e.g., A). So genotypes aa, Aa and AA are coded as
0, 1 and 2,
respectively. See Table~\ref{AMkernel} for the numerical genotype
coding of the three individuals at the ten marker position.

At a certain SNP position, the allele matching score of the genotypes
for two individuals $i$ and $j$ (denoted as AM$_{ij}$), is
calculated as the total number of alleles which are identical by
state (see Table~\ref{AM}).
%
%
\begin{table}
\tablewidth=190pt
\caption{Calculation of the allele matching score}\label{AM}
\begin{tabular*}{\tablewidth}{@{\extracolsep{\fill}}lccc@{}}
\hline
& \multicolumn{3}{c@{}}{$\bolds{g_{j,s}}$}\\[-4pt]
& \multicolumn{3}{c@{}}{\hrulefill}\\
$\bolds{g_{i,s}}$ & \multicolumn{1}{c}{\textbf{AA(2)}} & \multicolumn{1}{c}{\textbf{Aa(1)}}
& \multicolumn{1}{c@{}}{\textbf{aa(0)}}\\
\hline
AA(2) & 4 & 2 & 0\\
Aa(1) & 2 & 2 & 2\\
aa(0) & 0 & 2 & 4\\
\hline
\end{tabular*}
\end{table}
Then the genomic similarity score across the whole gene between two
individuals $i$ and $j$ (denoted as GSS$_{ij}$) is then calculated as
the weighted sum of allele matching scores for
all SNP markers in that gene (see Table~\ref{AMkernel}).

%
%
\begin{table}
\tabcolsep=0pt
\caption{Illustration on the calculation of genomic similarity
score}\label{AMkernel}
\begin{tabular*}{\tablewidth}{@{\extracolsep{4in minus 4in}}lcccccccccc@{}}
\hline
&\textbf{SNP}$\bolds{_1}$&
\textbf{SNP}$\bolds{_2}$&\textbf{SNP}$\bolds{_3}$
&\textbf{SNP}$\bolds{_4}$&\textbf{SNP}$\bolds{_5}$
&\textbf{SNP}$\bolds{_6}$&\textbf{SNP}$\bolds{_7}$
&\textbf{SNP}$\bolds{_8}$&\textbf{SNP}$\bolds{_9}$
& \textbf{SNP}$\bolds{_{10}}$\\
\hline
Individual 1 & 2 & 0 & 2 & 1 & 1 & 0 & 1 & 1 & 1 & 1\\
Individual 2 & 0 & 0 & 0 & 0 & 0 & 0 & 0 & 1 & 0 & 0\\
Individual 3 & 0 & 0 & 0 & 1 & 1 & 0 & 1 & 0 & 1 & 1\\
[4pt]
AM$_{12}$ & 0 & 4 & 0 & 2 & 2 & 4 & 2 & 2 & 2 & 2\\
GSS$_{12}$ & \multicolumn{10}{c@{}}{$(0+4+0+2+2+4+2+2+2+2)/(4\times
10)=0.5$}\\
AM$_{13}$ & 0 & 4 & 0 & 2 & 2 & 4 & 2 & 2 & 2 & 2\\
GSS$_{13}$ & \multicolumn{10}{c@{}}{$(0+4+0+2+2+4+2+2+2+2)/(4\times
10)=0.5$}\\
AM$_{23}$ & 4 & 4 & 4 & 2 & 2 & 4 & 2 & 2 & 2 & 2\\
GSS$_{23}$& \multicolumn{10}{c@{}}{$(4+4+4+2+2+4+2+2+2+2)/(4\times
10)=0.7$}\\
\hline
\end{tabular*}
\end{table}

With the genomic similarity measurements between all
possible individual pairs [in the example, $(1,2)$, $(1,3)$ and $(2,3)$],
the $3\times3$ kernel\vadjust{\goodbreak} matrix based on the example data is given as
\[
\mathbf{K}=\left[
\matrix{
1 & 0.5 & 0.5\cr
0.5 & 1 & 0.7\cr
0.5 & 0.7 & 1}
\right].
\]
\end{appendix}

\section*{Acknowledgments}

We thank Editor K. Lange, the Associate Editor and two anonymous
referees for their insightful comments that greatly improved the manuscript.
We also thank Dr. R. Romero for providing the birth weight data,
Professor J. Stapleton for his careful reading of the manuscript, and
Mr. B. Rosa and Dr. J. Chen for helpful discussion on using the Cytoscape
program. The computation of the work is supported by Revolution~R
(\url{http://www.revolutionanalytics.com/}).


%

\printaddresses

\end{document}